\documentclass[aps,12pt]{revtex4}
\usepackage{epsfig}
\usepackage{amsmath}
\usepackage{bm}
\usepackage{times}

\begin{document}
\title{Predictions from type II see-saw mechanism in $SU(5)$}
\author{Ilja Dor\v{s}ner}
\email{idorsner@phys.psu.edu}
\author{Irina Mocioiu}
\email{irina@phys.psu.edu} \affiliation{
The Pennsylvania State University \\
104 Davey Lab, University Park, PA 16802}
\begin{abstract}
We propose a simple, testable, $SU(5)$ model within the context of the
type II neutrino see-saw mechanism. It is based on requiring
renormalizability, the absence of any other matter fields besides
those already present in the Standard Model and consistency with all
experimental data. These ``minimal'' requirements, together with
group-theoretical considerations, uniquely determine the model and
lead to interesting implications. The model predicts correlation
between a light $SU(2)$ triplet boson responsible for the type II
see-saw mechanism and observable proton decay signatures. It also
allows for an enhanced production of doubly charged Higgs particles
through the $WW$ fusion process due to a built-in custodial
symmetry. This could also have profound impact on the explicit
realization of electroweak symmetry breaking. The model also predicts
the existence of a light scalar that transforms as a colour octet and
electroweak doublet, with interesting phenomenological consequences.
\end{abstract}
\maketitle

\section{Introduction}

Grand unified theories have been used for a long time as a very
elegant framework of physics beyond the Standard Model (SM). While
they are tightly constrained by limits on the proton decay
lifetime and by the requirement of gauge coupling unification,
grand unified models are typically quite complicated and hard to
test. The existence of non-zero neutrino masses and mixings has
brought new experimental evidence for physics beyond the Standard
Model. Understanding the neutrino properties in grand unified
theories comes rather naturally through the see-saw mechanism,
where integrating out large masses leads to the appearance of
small masses. There are three types of see-saw models that can
provide an understanding of the neutrino phenomenology: type I
see-saw models require the existence of SM singlets that have
Dirac Yukawa couplings to SM leptons; type II see-saw models use
an $SU(2)$ scalar triplet with Majorana type couplings to SM
leptons; type III models couple a fermionic $SU(2)$ triplet to SM
leptons though a Dirac Yukawa type coupling. While naturally
explaining many observed features, the main challenge when
building grand unified models and specific types of embedding the
see-saw mechanism within these is to find the means for testing
the validity of such models. Naturally, a lot of effort has been
directed towards building ``minimal'' models, where ``minimal''
has been understood in many different ways, but always with the
goal of introducing a small number of unknowns in order to keep
the theory predictive and testable. In this paper we discuss a
highly constrained, testable $SU(5)$ model and its possible
consequences. It is based on a small set of ``minimal''
requirements: renormalizability, the absence of any other matter
fields besides those already present in the Standard Model and
consistency with all present experimental data. Together with
group theoretical considerations, these requirements uniquely
determine the model and its implications.

We introduce the model in section \ref{scalars} and discuss how it
addresses the required symmetry breaking, generation of fermion
masses and mixing angles. The model has built-in a type II
mechanism for neutrino mass generation and it contains a number of
interesting scalars, including two electroweak doublets and two
electroweak triplets.

In section \ref{protondecay} we address in detail the issue of proton
decay due to gauge boson and scalar exchange, including partial
widths, experimental constraints and flavour dependence.

In section \ref{gaugecoupling} we discuss gauge coupling
unification. We show how the requirement of unification implies
the existence of light scalars, as well as observable proton decay
signatures.

In section \ref{electroweak} we start the discussion of the
phenomenological implications of the model. One of the very
interesting features of the model is that, while having a
complicated Higgs structure and many potentially light scalars, it
can preserve custodial symmetry at tree level, such that
electroweak constraints can be greatly relaxed compared to simple
generic models. We discuss some of the possible collider
signatures of the model, in particular the phenomenology of a
doubly charged scalar. This has been extensively studied in
various contexts and its dominant production mechanism is usually
the Drell-Yan process. We emphasize here that in our model it is
possible to have a large parameter space where $WW$ fusion becomes
the dominant production channel. We also discuss correlations
between potential collider signatures and proton decay
observations. In section \ref{octet} we discuss the potential
implications of a colour octet, electroweak doublet, which
appears in our model and is rather light.

In section \ref{compare} we present a comparison of our model with
other $SU(5)$ grand unified models.

We present our conclusions in section \ref{conclusions}.

\section{Scalar sector: symmetry breaking, fermion mass generation}
\label{scalars}

As previously mentioned, the $SU(5)$ model we investigate is the
simplest possible realization that satisfies the following
requirements: renormalizability, the absence of other matter fields
besides the ones that have already been observed experimentally and a
viable phenomenology. We should thus be able to generate both the
breaking of $SU(5)$ to the SM group as well as the SM symmetry
breaking, all fermion masses and their mixing angles and gauge
coupling unification, as well as be consistent with proton decay and
other experimental constraints.

The scalars present in the model are determined by the above
requirements. Let us start by specifying the representations
responsible for the fermion mass generation. The $i$th generation
of the SM matter fields resides in $\bm{10}_i$ and
$\overline{\bm{5}}_i$~\cite{Georgi:1974sy}. To be specific
$\bm{10}_i=(\bm{1},\bm{1},1)_i\oplus(\overline{\bm{3}},\bm{1},-2/3)_i
\oplus(\bm{3},\bm{2},1/6)_i=(e^C_i,u^C_i,Q_i)$ and
$\overline{\bm{5}}_i=(\bm{1},\bm{2},-1/2)_i\oplus
(\overline{\bm{3}},\bm{1},1/3)_i=(L_i,d^C_i)$, where $Q=(u \qquad
d)^T$, $L=(\nu \qquad e)^T$ and $i=1,2,3$. In order to generate
mass through a renormalizable operator, a Higgs representation
must have a component that is both electrically neutral and
$SU(3)$ colour singlet and be in the tensor product of the
appropriate matter field representations. For the up quarks one
needs either a $\bm{5}$ or $\bm{45}$ of Higgs since the up quark
mass originates from the contraction of $\bm{10}_i$ and
$\bm{10}_j$: $\bm{10} \times \bm{10} =
\overline{\bm{5}}\oplus\overline{\bm{45}}\oplus\overline{\bm{50}}$.
Only $5$ and $45$ dimensional representations have a component
that is both electrically neutral and colour singlet that could
thus obtain a phenomenologically allowed vacuum expectation value
(VEV). The down quark and charged lepton masses originate from the
contraction of $\bm{10}_i$ with $\overline{\bm{5}}_j$: $\bm{10}
\times \overline{\bm{5}} = \bm{5}\oplus\bm{45}$. This time both
the $\bm{5}$ and $\bm{45}$ of Higgs are needed to obtain
phenomenologically allowed masses. Neutrinos on the other hand
reside in $\overline{\bm{5}}_i$. Their Majorana mass originates
from the symmetric contraction of $\overline{\bm{5}}_i$ with
$\overline{\bm{5}}_j$. Recall, $\overline{\bm{5}} \times
\overline{\bm{5}} = \overline{\bm{10}} \oplus \overline{\bm{15}}$,
where $\overline{\bm{15}}$ ($\overline{\bm{10}}$) is a
(anti)symmetric representation. Hence, to generate the Majorana
neutrino masses at the tree level, one must use a $\bm{15}$ of
Higgs which happens to have a neutral component as part of a $Y=2$
$SU(2)$ triplet. This is an $SU(5)$ implementation of the
so-called type II see-saw
mechanism~\cite{Schechter:1980gr,Lazarides:1980nt,Mohapatra:1980yp}.
In addition to $5$, $15$ and $45$ dimensional scalar
representations one also needs a $\bm{24}$ of Higgs in order to
break the $SU(5)$ symmetry. These representations decompose under
the SM as $\bm{5}= (\Psi_D, \Psi_T) =
(\bm{1},\bm{2},1/2)\oplus(\bm{3},\bm{1},-1/3),$ $ \bm{15}=
(\Phi_a, \Phi_b, \Phi_c) = (\bm{1},\bm{3},1)\oplus
(\bm{3},\bm{2},1/6)\oplus(\bm{6},\bm{1},-2/3)$, $\bm{24}=(\Sigma_8
, \Sigma_3, \Sigma_{(3,2)}, \Sigma_{(\overline{3},2)},
\Sigma_{24}) = (\bm{8},\bm{1},0)\oplus(\bm{1},\bm{3},0)
\oplus(\bm{3},\bm{2},-5/6)\oplus(\overline{\bm{3}},\bm{2},5/6)
\oplus(\bm{1},\bm{1},0)$, $\bm{45}=(\Delta_1, \Delta_2, \Delta_3,
\Delta_4, \Delta_5, \Delta_6, \Delta_7) =
(\bm{8},\bm{2},1/2)\oplus (\overline{\bm{6}},\bm{1}, -1/3) \oplus
(\bm{3},\bm{3},-1/3) \oplus (\overline{\bm{3}}, \bm{2}, -7/6)
\oplus (\bm{3},\bm{1}, -1/3) \oplus (\overline{\bm{3}}, \bm{1},
4/3) \oplus (\bm{1}, \bm{2}, 1/2)$. This decomposition will be
useful when we discuss gauge coupling unification and proton
decay. This completes the specification of the Higgs sector that
is uniquely determined by group-theoretical considerations once
our requirements are imposed. In what follows we will always
assume that all terms allowed by the gauge symmetry are present in
the Lagrangian density and specify them only when necessary.

A beautiful feature of $SU(5)$ is that the phenomenologically
allowed symmetry breaking chain is unique, i.e, $SU(5) \rightarrow
SU(3) \times SU(2) \times U(1) \rightarrow SU(3) \times
U(1)_{\textrm{em}}$. The grand unified symmetry is broken down to
the SM by the VEV of the SM singlet $\Sigma_{24}$ in the $\bm{24}$
of Higgs: $\langle\bm{24}\rangle = v_{24}/\sqrt{30} \,
\textrm{diag}(2,2,2,-3,-3)$. The symmetry breaking takes place at
the so-called GUT scale $M_{GUT}$ where the SM gauge couplings
unify into $g_{GUT}$. At this stage proton decay mediating gauge
bosons $X$ ($Y$) absorb the $\Sigma_{(3,2)}$
($\Sigma_{(\overline{3},2)}$) degrees of freedom and become
massive. Their masses are ($M_X \cong M_Y \equiv
)M_V=M_{GUT}=\sqrt{\frac{5}{12}} v_{24} g_{GUT}$. Due to this
particular feature one is able to make accurate statements with
regard to proton decay signatures via gauge mediation. The
electroweak symmetry of the SM is subsequently broken by the VEVs
of $SU(2)$ doublets $\Psi_D$ and $\Delta_7$ as well as the VEVs of
$SU(2)$ triplets $\Phi_a$ and $\Sigma_3$. The first two are the
sources of the charged fermion masses while the third one is the
generator of neutrino masses. The VEV of $\Sigma_3$, on the other
hand, affects masses of $X$, $Y$, $W$ and $Z$ gauge bosons. We
will see that the fields that participate in electroweak symmetry
breaking should be light in order to have a phenomenologically
viable model.

Fermion masses follow from the Yukawa potential:
\begin{eqnarray}
V &=& (Y_1)_{ij} \ (\bm{10}^{\alpha \beta})_i
(\overline{\bm{5}}_{\alpha})_j \bm{5}^*_\beta + (Y_2)_{ij} \
(\bm{10}^{\alpha \beta})_i (\overline{\bm{5}}_{\delta})_j
\bm{45}^{*\delta}_{\alpha \beta} +(Y_3)_{ij} \
(\overline{\bm{5}}_{\alpha})_i
 (\overline{\bm{5}}_{\beta})_j \bm{15}^{\alpha \beta} \nonumber\\
&+&\epsilon_{\alpha \beta \gamma \delta \epsilon} \left[
(Y_4)_{ij} \ (\bm{10}^{\alpha \beta})_i (\bm{10}^{\gamma
\delta})_j \bm{5}^\epsilon + (Y_5)_{ij} \ (\bm{10}^{\alpha
\beta})_i (\bm{10}^{\zeta \gamma})_j (\bm{45})^{\delta
\epsilon}_\zeta \right],\qquad i,j=1,2,3,
\end{eqnarray}
where Greek indices are contracted in the $SU(5)$ space.
The mass matrices, in an obvious notation, are
\begin{subequations}
\label{M}
\begin{eqnarray}
\label{M_D}
M_D &=& \left( Y^T_1 \ v_5^* \ + \ 2 \ Y^T_2 \ v_{45}^* \right)/\sqrt{2} ,\\
M_E &=& \left( Y_1 \ v_5^* \ - 6 \ Y_2 \ v_{45}^*
\right)/\sqrt{2},
\label{GJ}\\
M_N &=& Y_3 \ v_{15},\\
\label{M_U} M_U &=& \left[ 4 \ (Y^T_4+Y_4) \ v_5 \ -  \ 8 \
(Y^T_5-Y_5) \ v_{45}\right]/\sqrt{2},
\end{eqnarray}
\end{subequations}
where $\langle\bm{5}\rangle=v_5/\sqrt{2}$,
$\langle\bm{45}\rangle^{1 5}_{1}= \langle\bm{45}\rangle^{2
5}_{2}=\langle\bm{45}\rangle^{3 5}_{3} =v_{45}/\sqrt{2}$ and
$\langle\bm{15}\rangle=v_{15}$. $Y_1$, $Y_2$, $Y_4$ and $Y_5$ are
arbitrary $3 \times 3$ Yukawa matrices, while $Y_3$ represents a
symmetric $3 \times 3$ matrix. The factor of 3 difference between
the second terms in Eqs.~\eqref{M_D} and~\eqref{GJ} is the
so-called Georgi-Jarlskog~\cite{Georgi:1979df} factor. Its origin
is due to the fact that $\bm{45}$ satisfies the following
conditions: $(\bm{45})^{\alpha \beta}_{\delta} = -
(\bm{45})^{\beta \alpha}_{\delta}$, $\sum_{\alpha=1}^5
(\bm{45})^{\alpha \beta}_{\alpha} = 0$. Hence, one has
$\sum_{i=1}^3 \langle\bm{45}\rangle^{i 5}_{i} = - \langle\bm{45}
\rangle^{4 5}_4$. The fermion mass eigenstate basis is defined
through the following transformations: $ U^T_C \ M_U \ U =
M_U^{\textrm{diag}}$, $D^T_C \ M_D \ D = M_D^{\textrm{diag}}$,
$E^T_C \ M_E \ E = M_E^{\textrm{diag}}$ and $N^T \ M_N \ N =
M_N^{\textrm{diag}}$. $M_{U,D,E,N}^{\textrm{diag}}$ represent
diagonal matrices with real eigenvalues.

Eqs.~\eqref{M_D} and~\eqref{GJ} imply $M_E^T=(-3) M_D$ if a $5$
($45$) dimensional Higgs representation is present. In other
words, the one Higgs doublet scenario predicts
$m_\tau/m_b=m_\mu/m_s=m_e/m_d$ at the GUT scale, which is in
conflict with experimental observations. This is why both the
$\bm{5}$ and $\bm{45}$ of Higgs are needed. Eq.~\eqref{M_U}, on
the other hand, shows that $\bm{5}$ ($\bm{45}$) induces a
symmetric (antisymmetric) part in $M_U$. This is very important
for the discussion of the flavour dependence of the proton decay
signatures. Only if $M_U$ is a purely symmetric matrix does this
dependence disappear in some of the decay
channels~\cite{FileviezPerez:2004hn}. It is thus clear that in
\textit{any}\/ realistic model the $SU(5)$ symmetry cannot insure
even a partial absence of the flavour dependence in the proton
decay
signatures~\cite{DeRujula:1980qc,FileviezPerez:2004hn,Dorsner:2004xx,Dorsner:2004xa}.
We will address this issue in more detail in section
\ref{protondecay}.

Although the Higgs sector looks rather cumbersome, it is the
simplest one that yields satisfactory phenomenology while
preserving the matter content of the SM. At this point it seems
difficult for the model to have any firm and testable predictions
unless some additional assumptions are imposed. Fortunately, as we
soon demonstrate, the model does predict experimentally observable
proton decay. It also predicts that some of the scalars have to be
light enough to be of experimental interest in order for
unification to take place. Here we refer to $\Psi_D$, $\Phi_a$,
$\Sigma_3$, $\Delta_1$ and $\Delta_7$. If some of these fields are
not very light they would jeopardize proton stability and hence
rule out the model. In addition, there is a clear correlation
between a light $\Phi_a$ and the proton decay signatures that
could allow unambiguous determination of the underlying mechanism
of the neutrino mass generation. Recall, the VEV of $\Phi_a$
generates massive neutrinos. We thus turn to the discussion of
proton decay signatures and constraints.

\section{Proton decay}
\label{protondecay}
Our main predictions rely strongly on consistent application of
the current experimental bounds on the partial proton decay
lifetimes that constrain the mass spectrum of the model.

The vector gauge boson $d=6$ operators contributing to the decay
of the proton in the $SU(5)$ framework are
well-known~\cite{Weinberg:1979sa,Weinberg:1980bf,Weinberg:1981wj,Wilczek:1979hc,Sakai:1981pk}:
\begin{subequations}
\label{SU5}
\begin{eqnarray}
\label{O1} \mathcal{O}_1&=& k^2 \ \epsilon_{ijk} \
\epsilon_{\alpha \beta} \ \overline{u_{i a}^C} \ \gamma^{\mu} \
Q_{j \alpha a}   \
\overline{e_b^C} \ \gamma_{\mu} \ Q_{k \beta b},\\
\label{O2} \mathcal{O}_2&=& k^2 \ \epsilon_{ijk} \
\epsilon_{\alpha \beta} \ \overline{u_{i a}^C} \ \gamma^{\mu} \
Q_{j \alpha a}   \ \overline{d^C_{k b}} \ \gamma_{\mu} \ L_{\beta
b}.
\end{eqnarray}
\end{subequations}
$i$, $j$ and $k$ are the colour indices, $a$ and $b$ are the family
indices, $\alpha, \beta =1,2$ and $k^2= 2 \pi \alpha_{GUT}
M^{-2}_{(X,Y)}$.

The effective operators for decay channels take the following form
in the physical basis~\cite{FileviezPerez:2004hn}:
\begin{subequations}
\begin{eqnarray}
\label{1Oec1} \mathcal{O}(e_{\alpha}^C, d_{\beta})&=& k^2 \left[
V^{11}_1 V^{\alpha \beta}_2 + ( V_1 V_{UD})^{1 \beta}( V_2
V^{\dagger}_{UD})^{\alpha 1}\right] \ \epsilon_{ijk}
\overline{u^C_i}  \gamma^{\mu}  u_j  \overline{e^C_{\alpha}}
\gamma_{\mu}  d_{k \beta}, \\
\label{1Oe} \mathcal{O}(e_{\alpha}, d^C_{\beta})&=& k^2 V^{11}_1
V^{\beta \alpha}_3 \ \epsilon_{ijk}  \overline{u^C_i} \gamma^{\mu}
u_j  \overline{d^C_{k \beta}}  \gamma_{\mu} \ e_{\alpha},\\
\label{1On} \mathcal{O}(\nu_l, d_{\alpha}, d^C_{\beta} )&=& k^2 (
V_1 V_{UD} )^{1 \alpha} ( V_3 V_{EN})^{\beta l} \ \epsilon_{ijk}
 \overline{u^C_i}  \gamma^{\mu}  d_{j \alpha}
\overline{d^C_{k \beta}}  \gamma_{\mu}  \nu_l, \qquad l=1,2,3.
\end{eqnarray}
\end{subequations}
$V_1= U_C^{\dagger} U$, $V_2=E_C^{\dagger}D$,
$V_3=D_C^{\dagger}E$, $V_{UD}=U^{\dagger}D=K_1 V_{CKM} K_2$ and
$V_{EN}=E^{\dagger}N=K_3 V_{PMNS}$ are unitary mixing matrices.
$K_{1,\,3}$ and $K_2$ are diagonal matrices containing three and
two phases, respectively. $V_{CKM}$ ($V_{PMNS}$) is the usual
Cabibbo-Kobayashi-Maskawa (Pontecorvo-Maki-Nakagawa-Sakata) matrix
that describes the mixing angles and phases of quarks (leptons).

In what follows we will focus our attention on the proton decay
into either a $\pi$ or $K$ meson and charged antilepton. For a
discussion that treats decays into antineutrinos
see~\cite{PavelNathReview}. The widths for the decays into charged
antileptons are:
\begin{eqnarray*}
\label{ratio3} \Gamma (p \rightarrow \pi^0
e_{\beta}^+)&=&\frac{C(p,\pi)}{2}  A_1^2 \left[A^2_{S\,R}
\left|V^{11}_1 V_3^{1\beta} \right|^2+A^2_{S\,L} \left|V^{11}_1
V_2^{\beta 1}+(V_1 V_{UD})^{11} (V_2 V^\dagger_{UD})^{\beta 1} \right|^2 \right],\\
\label{ratio4} \Gamma (p \rightarrow K^0 e_{\beta}^+)&=& C(p,K)
A_2^2 \left[ A^2_{S\,R} \left|V^{11}_1 V_3^{2\beta}
\right|^2+A^2_{S\,L} \left|V^{11}_1 V_2^{\beta 2}+(V_1
V_{UD})^{12} (V_2 V^\dagger_{UD})^{\beta 1} \right|^2 \right],
\end{eqnarray*}
where
\begin{equation}
C(a,b)= \frac{(m_a^2 - m_b^2)^2}{8 \pi m_a^3 f^2_{\pi}} \ A_L^2 \
|\alpha|^2 \ k^4.
\end{equation}

The relevant $A_i$ factors are: $A_1= 1 + D+F$ and $A_2 =1 +
\frac{m_p}{m_B}(D - F)$~\cite{PavelNathReview}. To generate
numerical results we use $m_p=938.3$\,MeV, $D=0.81$, $F=0.44$,
$m_B=1150$\,MeV, $f_{\pi}=139$\,MeV, $A_L=1.25$,
$|V_{ud}|=0.97377$, $|V_{ub}|=3.96 \times 10^{-3}$, and
$\alpha=0.015$\,GeV$^3$~\cite{Aoki:2004xe}. Here, $\alpha$ is the
so-called matrix element. In addition one needs to evaluate the
leading-log renormalization of the operators
$\mathcal{O}(e_{\alpha}^C, d_{\beta})$ and
$\mathcal{O}(e_{\alpha}, d^C_{\beta})$ from the GUT scale to $M_Z$
which is described by the coefficients $A_{SL}$ and $A_{SR}$
respectively. (The QCD running below $M_Z$ is captured by the
coefficient $A_L$.) These coefficients
are~\cite{Buras:1977yy,Ellis:1979hy,Wilczek:1979hc}:
\begin{equation}
A_{S\,L(R)}=\prod_{i=1,2,3} \prod_I^{M_Z \leq M_I \leq M_{GUT}}
\left[\frac{\alpha_i(M_{I+1})}{\alpha_i(M_I)}\right]^{\frac{\gamma_{L(R)i}}{\sum_J^{M_Z
\leq M_J \leq M_I} b^J_i}},\,\, \gamma_{L(R)i}=(23(11)/20,9/4,2).
\end{equation}
$b_{i}^J$ are the usual $\beta$-function coefficients due to
particle $J$ of mass $M_J$ and $\alpha_i(M_I)$ are the gauge
coupling constants at the scale $M_I$. We use the following
experimental values at $M_Z$ in the $\overline{MS}$
scheme~\cite{PDG}: $\alpha_3 = 0.1176 \pm 0.0020$, $\alpha^{-1} =
127.906 \pm 0.019$ and $\sin^2 \theta_W = 0.23122 \pm 0.00015$.

To predict the partial lifetimes of the proton for these decay
channels we still need to know $k$, $V^{1b}_1$, $V_2$ and $V_3$.
In addition there are two diagonal matrices containing CP
violating phases, $K_1$ and $K_2$. Therefore it is impossible to
test a general $SU(5)$ scenario through the decay of the proton
unless we specify both the full flavour structure and mass
spectrum of the GUT model. What is then usually assumed for the
flavour structure, in order to extract a conservative limit on the
GUT scale, is that $U_C = U$, $D_C=E$ and $E_C = D$. Under these
assumptions the dominant proton decay mode is $p \rightarrow \pi^0
e^+$ and the theoretical prediction for this channel comes out to
be $\tau^{\textrm{theo.}}=3.1 \times 10^{33}
(M_{GUT}/10^{16}\,\textrm{GeV})^4 \alpha_{GUT}^{-2}
(\alpha/0.015\,\textrm{GeV}^3)(A_{S\,R}^2+3.8
A_{S\,L}^2)^{-1}$\,years. The current experimental limit on the
partial lifetime $\tau^{\textrm{exp.}}>4.4 \times
10^{33}$\,years~\cite{Ganezer:2001qk} thus translates into the
following bound on $M_{GUT}$: $M_{GUT}> 2.6 \times 10^{16}
\sqrt{\alpha_{GUT}}\,\textrm{GeV}$ where we take
$A_{S\,L}=A_{S\,R}=2.5$. Of course, if both the particle content
and mass spectrum of the model are known it is possible to
evaluate $A_{S\,L}$ and $A_{S\,R}$ more accurately.

For the proton decay through scalar exchange---for example via
$\Psi_T$---the relevant couplings are Yukawa couplings of the
first and second generation that are expected to be of the order
of $Y \simeq 10^{-6}$--$10^{-4}$. We thus get the relevant scale
at which scalar exchange becomes dominant by replacing
$\alpha_{GUT}$ with $Y^2$. This in turn yields a lower bound on
the phenomenologically allowed scalar mass to be around
$10^{12}$\,GeV.

Finally, let us for completeness discuss the flavour dependence of
the experimental bound on $M_{GUT}$. We will assume for the sake
of argument the following flavour scenario~\cite{Dorsner:2004xa}:
$(V_1 V_{UD})^{1\alpha}=0$ and $V_2^{\alpha\beta}=
V_3^{\alpha\beta}=0$ ($\alpha=1$ or $\beta=1$).
It is easy to see from Eq.~\eqref{1On} that there will be no
decays into antineutrinos while the only surviving channel with an
antilepton in the final state is actually $p \rightarrow K^0
\mu^+$. We get $\Gamma (p \rightarrow K^0 \mu^+)= C(p,K) A_2^2
\left[A^{2}_{SR} +A^{2}_{SL} \right] |V_{ub}|^2$ which is more
then six orders of magnitude smaller than the decay width when
$U_C = U$, $D_C=E$ and $E_C = D$. Both flavour scenarios are a
priori possible.

In order to show that our model predicts observable proton decay
signatures and prefers if not predicts certain light scalars
including the $SU(2)$ triplets with $Y=0$ and $Y=2$, we need to
address the issue of gauge coupling unification with these results
in mind. We will explicitly assume the flavour scenario where $U_C
\approx U$, $D_C \approx E$ and $E_C \approx D$ in what follows.

\section{Gauge coupling unification}
\label{gaugecoupling}
The behavior of the gauge couplings between the electroweak and
the GUT scale is described by three renormalization group
equations---one for each gauge coupling of the SM $\alpha_i$
($i=1,2,3$). If we impose unification and accordingly eliminate
the unified coupling constant $\alpha_{GUT}$, we are left with only
two relevant equations~\cite{Giveon:1991zm}. These are:
\begin{subequations}
\begin{eqnarray}
\label{condition1} \frac{B_{23}}{B_{12}}&=&\frac{5}{8}
\frac{\sin^2
\theta_W-\alpha/\alpha_3}{3/8-\sin^2 \theta_W}=0.716 \pm 0.005,\\
\label{condition2} \ln \frac{M_{GUT}}{M_Z}&=&\frac{16 \pi}{5
\alpha} \frac{3/8-\sin^2 \theta_W}{B_{12}}=\frac{184.9 \pm
0.2}{B_{12}},
\end{eqnarray}
\end{subequations}
where the right-hand sides reflect the latest experimental
measurements of the SM parameters~\cite{PDG}.

In view of the fact that we are interested in proton decay
signatures of our model, Eq.~\eqref{condition2} is especially
interesting. Namely, for a given minimal $B_{12}$ value of a
specific model, it is possible to obtain associated upper bound on
$M_{GUT}$, which is a crucial ingredient for accurate proton decay
predictions. The $B_{ij}$ coefficients on the other hand depend on
the specific particle spectrum. More precisely, $B_{ij}=B_i -
B_j$, where $B_i$ coefficients are given by:
\begin{equation}
\label{r} B_i = \sum_{I} b_{i}^I r_{I}, \qquad r_I=\frac{\ln
M_{GUT}/M_{I}}{\ln M_{GUT}/M_{Z}}, \qquad (0 \leq r_I \leq 1).
\end{equation}
The SM content with $n$ light Higgs doublet fields has
$B_1=40/10+n/10$, $B_2=-20/6+n/6$ and $B_3=-7$. Hence the SM case
($n=1$) yields $B^{\textrm{SM}}_{23}/B^{\textrm{SM}}_{12}=0.53$.
Clearly, additional particles with masses below the GUT scale are
required for successful unification. In addition to satisfying
Eq.~\eqref{condition1}, any potentially realistic grand unified
scenario must generate large enough GUT scale in order to satisfy
the proton decay constraints. The careful analysis of the $X$ and
$Y$ mediated proton decay from the previous section implies a
lower bound on the GUT scale in $SU(5)$ to be $M_{GUT}> 4$--$5
\times 10^{15}$\,GeV. Again, we have assumed that $U_C \approx U$,
$D_C \approx E$ and $E_C \approx D$.

Our first aim is to show that successful unification implies
proton decay signatures that are within the reach of the future
proton decay experiments regardless of the exact mass spectrum of
the scalars in our model. Our discussion will also imply that some
of the scalars---those that do not mediate proton decay---are
\textit{always}\/ very light in order to have phenomenologically
acceptable proton decay widths. We thus start by looking at the
impact of light scalars $I$ with negative contributions toward the
$B_{12}$ coefficient ( i.e., $\Delta b_{12}^I=b_{1}^I-b_{2}^I<0$)
on unification, since only those fields can raise the GUT scale.
We show that, if the vector gauge boson mediated proton decay is
suppressed beyond the experimentally established limit, then the
proton decay due to scalar exchange is experimentally accessible
and vice versa.

The multiplets with negative contribution to $B_{12}$ are
$\Phi_D$, $\Sigma_3$, $\Delta_1$, $\Delta_3$, $\Delta_7$, $\Phi_a$
and $\Phi_b$. We have underlined them for convenience in
Table~\ref{tab:table1} where we list all the $\Delta
b_{ij}=b_i-b_j$ contributions. $\Delta_3$ and $\Phi_b$ cannot be
arbitrarily light in order to avoid existing experimental
limits on partial proton decay lifetimes. The fact that a
$\Delta_3$ exchange could contribute to proton decay has been recently
pointed out~\cite{Dorsner:2006dj}. Other fields that
mediate proton decay but have positive $B_{12}$ contributions are
$\Psi_T$, $\Delta_5$ and $\Delta_6$. We have placed a line over
them in Table~\ref{tab:table1} for convenience. All these scalars
should have masses of the order of $10^{12}$\,GeV or higher unless
some special arrangements take place in the Yukawa sector that
would suppress their contributions to proton decay. 
The important point is that if these
scalar fields are as light as $10^{12}$\,GeV their proton decay
signatures would be at their present experimentally established
limits. With that in mind we now determine an upper bound on the
GUT scale.

We take the fields that do not mediate proton decay and set their
masses to $M_Z$, i.e.,
$r_{(\Psi_D,\Sigma_3,\Delta_1,\Delta_7,\Phi_a)}=1$, to get
$B_{12}=(110/15-1/15-15/15)$. The first two contributions are the
usual SM contributions to $B_{12}$ while the net effect of all
other scalar multiplets on $B_{12}$ is rather small. In fact, this
yields via Eq.~\eqref{condition2} that $M_{GUT} \simeq 6 \times
10^{14}$\,GeV. This is clearly below the lower bound on $M_{GUT}$
as inferred from the experimentally measured limits on proton
decay lifetime. If we now take into account that $\Delta_3$ and
$\Phi_b$ could only be as light as $10^{12}$\,GeV, which roughly
translates into $r_{(\Delta_3,\Phi_b)}\leq 1/3 $, we obtain
$M_{GUT} \leq 3 \times 10^{16}$\,GeV which should be considered as
a conservative upper bound on the GUT scale in our model at
one-loop. If any of the fields with negative contributions to
$B_{12}$ is actually heavier than what we have assumed, then the
GUT scale would accordingly go down in proportion to the
corresponding $b_{12}$ contribution. In particular, if we want to
suppress proton decay rates due to the scalar exchange by taking
$\Delta_3$ and $\Phi_b$ masses to be significantly above
$10^{12}$\,GeV, we would significantly enhance decay rates due to
the vector gauge boson exchange.
\begin{table}[ht]
\caption{\label{tab:table1} The scalar $B_{ij}$ coefficient
contributions.}
\begin{ruledtabular}
\begin{tabular}{lcccccccccccccccc}
 & $\underline{\Psi_D}$  & $\overline{\Psi_T}$ & $\Sigma_8$
 & $\underline{\Sigma_3}$ & $\underline{\Delta_1}$ & $\Delta_2$
 & $\overline{\underline{\Delta_3}}$ & $\Delta_4$ & $\overline{\Delta_5}$
 & $\overline{\Delta_6}$ & $\underline{\Delta_7}$&
 $\underline{\Phi_a}$ & $\overline{\underline{\Phi_b}}$ & $\Phi_c$\\
\hline $\Delta b_{23}$ & $\frac{1}{6}$ & $-\frac{1}{6}$ &
$-\frac{3}{6}$ & $\frac{2}{6}$ & $-\frac{4}{6}$ & $-\frac{5}{6}$ &
$\frac{9}{6}$ & $\frac{1}{6}$ & $-\frac{1}{6}$ & $-\frac{1}{6}$ &
$\frac{1}{6}$&$\frac{4}{6}$
&$\frac{1}{6}$ &$-\frac{5}{6}$\\
$\Delta b_{12}$ &$-\frac{1}{15}$ & $\frac{1}{15}$ & 0 &
$-\frac{5}{15}$ & $-\frac{8}{15}$ &$\frac{2}{15}$ &
$-\frac{27}{15}$ & $\frac{17}{15}$ & $\frac{1}{15}$
& $\frac{16}{15}$ &  $-\frac{1}{15}$&$-\frac{1}{15}$ &$-\frac{7}{15}$ &$\frac{8}{15}$\\
\end{tabular}
\end{ruledtabular}
\end{table}

These are obviously good news as far as the testability of the
model is concerned. The model certainly implies that proton decay
should take place within the experimentally accessible
range~\cite{Jung:1999jq,Autiero:2007zj} regardless of the exact
scalar mass spectrum. In addition, some of the scalars with
negative $B_{12}$ contributions that do not mediate proton decay
are rather light. Here, in particular, we refer to $\Psi_D$,
$\Sigma_3$, $\Delta_1$, $\Delta_7$ and $\Phi_a$. $\Psi_D$ and
$\Delta_7$ are $SU(2)$ doublets, $\Sigma_3$ and $\Phi_a$ are
$SU(2)$ triplets while $\Delta_1$ transforms as a doublet of
$SU(2)$ and octet of $SU(3)$. Each of these fields is interesting
in its own right, especially from the point of view of accelerator
physics, which is an exciting prospect.

We have so far neglected the fact that Eqs.~\eqref{condition1}
and~\eqref{condition2} should be solved simultaneously. Let us do
that within the following scenario. Let us (i) fix the mass of
$\Phi_a$ to $300$\,GeV which will certainly be within the reach of
accelerator experiments, (ii) impose $M_{\Sigma_8} \geq 10^5$\,GeV
as required by nucleosynthesis considerations (see for example
discussion in Ref.~\cite{Bajc:2006ia} and references therein),
(iii) set $M_{\Delta_3}=M_{\Phi_b}=10^{12}$\,GeV and (iv) vary all
other fields in the model within their allowed range in order to
maximize $M_{GUT}$ via Eqs.~\eqref{condition2} and
Eq.~\eqref{condition1}. This simple exercise yields $M_{GUT} \leq
1.4 \times 10^{16}$\,GeV. This is in a good agreement with our
previous analysis. This value is obtained when
$\alpha_{GUT}^{-1}=29.4$, $M_{\Sigma_3}=M_Z$,
$M_{\Sigma_8}=10^5$\,GeV, $M_{\Delta_1}=M_Z$, $M_{\Delta_2}=2
\times 10^{10}$\,GeV and $M_{\Delta_7}=M_Z$. All other fields are
at the GUT scale. In this case the predicted proton lifetime for
$p \rightarrow \pi^0 e^+$ due to gauge mediation is a factor of
$51$ above the current experimental limit while the proton
lifetime due to scalar mediation is at the present limit.

If the scalar exchange induced proton decay is suppressed then the
vector boson exchange contributions is experimentally accessible.
To illustrate that we set $M_{\Delta_3}=M_{\Phi_b}=10^{13}$\,GeV
and keep $\Phi_a$ again at $300$\,GeV to obtain $M_{GUT} \leq 5.2
\times 10^{15}$\,GeV. The predicted proton lifetime through the
gauge boson mediation is then exactly at the current experimental
limit.

Note that in both cases we set some of the fields at the $M_Z$
scale, which is likely not realistic. In other words, the upper
bound on the GUT scale we discuss here is very conservative. With
this in mind we turn to the important question of testing this
model in collider 
experiments.

\section{Electroweak symmetry breaking sector}
\label{electroweak}
Let us start with the discussion of the electroweak symmetry breaking
sector of the model. It comprises two $SU(2)$ doublets---$\Psi_D$ and
$\Delta_7$---as well as two $SU(2)$ triplets---$\Phi_a$ and
$\Sigma_3$. Interestingly enough, this particular Higgs content can
preserve the custodial symmetry of the SM at tree level and thus
accommodate precision electroweak constraints. In fact, it corresponds to
the content of the models that have been tailor-made to accomplish
just that~\cite{Georgi:1985nv,Chanowitz:1985ug,Chivukula:1986sp}. Here
we have an example where the same kind of setup could naturally emerge
within a well-motivated GUT framework.

Since our primary concern is the possibility to test the underlying
see-saw mechanism, we observe that one of the consequences of this
custodial symmetry could be that the couplings of the see-saw triplet
$\Phi_a$ to the gauge bosons are much larger than expected from the
standard limits set by electroweak precision measurements.

Testing the electroweak symmetry breaking sector will be very
challenging due to its complexity. However, one nice
feature is that $\Phi_a$ contains a doubly charged Higgs boson
$\Phi_a^{\pm\pm}$ that does not mix with any other Higgs field in
the model. This makes the analysis of its experimental signatures
relatively model independent. With this in mind we limit our
discussion of accelerator signatures of light scalar particles
mainly to the $\Phi_a^{\pm\pm}$ production and subsequent decay.

There are a number of well-motivated models that all have
potentially light $Y=2$ triplet(s). These are primarily the
left-right symmetry
models~\cite{Senjanovic:1975rk,Mohapatra:1974gc}, little Higgs
models~\cite{Arkani-Hamed:2001nc,Arkani-Hamed:2002qx,Schmaltz:2004de},
and certain type II see-saw extensions of the SM~\cite{Ma:2000xh}.
Due to this and the fact that $\Phi_a^{\pm\pm}$ does not mix with
other Higgs fields there exists a large body of work on the doubly
charged Higgs signatures in
current~\cite{Acosta:2004uj,Akeroyd:2005gt,Abulencia:2007rd} and
future
colliders~\cite{Gunion:1989ci,Azuelos:2004dm,Akeroyd:2005gt,Hektor:2007uu,Han:2007bk}.
We accordingly point out only those salient features that could
make our model different from other models.

The dominant production of $\Phi_a^{\pm\pm}$s at the Tevatron and
LHC is either through the Drell-Yan (DY) $\Phi_a^{++}\Phi_a^{--}$
pair production or $WW$ fusion into a single doubly charged
component of $\Phi_a$. $WW$ fusion is proportional to the triplet
VEV which is primarily bounded from above by the electroweak
precision measurements due to its impact on the so-called $\rho$
parameter.  This bound is around $2$\,GeV within the framework of
the SM extended with a $Y=2$ triplet only. In our model however,
both $Y=0$ and $Y=2$ triplets get VEVs in addition to the two
Higgs doublets, i.e., $\langle\bm{5}\rangle=v_5/\sqrt{2}$,
$\langle\bm{45}\rangle^{1 5}_{1}= \langle\bm{45}\rangle^{2
5}_{2}=\langle\bm{45}\rangle^{3 5}_{3} =v_{45}/\sqrt{2}$, $\langle
\Sigma_3 \rangle=v'$ and $\langle\bm{15}\rangle=v_{15}$. If we
take all these VEVs into account the net tree-level contribution
to the $\rho$ parameter is
\begin{equation}
\rho=\frac{v_{5}^2+v_{45}^2+4 v_{15}^2+4
v'^{2}}{v_{5}^2+v_{45}^2+8 v_{15}^2}.
\end{equation}
The $W$ mass is given as $M_W=g^2/4 (v_{5}^2+v_{45}^2+4 v_{L}^2+4
v'^{2})$, where $\alpha_2=g^2/(4 \pi)$. $\sin^2 \theta_W$, on the
other hand, is not affected by additional VEVs at all. It is easy
to see that $\rho \simeq 1$ naturally at the tree level as long as
$v_{L} \simeq v'$, regardless of their absolute value. In fact,
$v_{L}( \simeq v')$ could be as large as $80$\,GeV as far as the
$\rho$ parameter and perturbativity of the top Yukawa constraints
are concerned at the tree level. This is possible in any $SU(5)$
scenario with~\cite{Joaquim:2006mn} or without
supersymmetry~\cite{Dorsner:2005fq,Dorsner:2005ii,Dorsner:2006hw}
that implements the type II see-saw mechanism, as well as in the
corresponding $SO(10)$ models.

There are thus two distinct regions in the parameter space of our
model in terms of $v_{L}$ values. If $v_{L} \simeq v' \simeq v_5
\simeq v_{45}$ then the $WW$ fusion into a doubly charged
component of $\Phi_a$ would overcome DY production of the
$\Phi_a^{++} \Phi_a^{--}$ pair and its subsequent decay could
primarily be into a $WW$ pair instead of a pair of charged
leptons. The $WW$ pair would eventually decay into a pair of
charged leptons and pair of neutrinos $10 \%$ of the time that
would then enable the detection of $\Phi_a^{++}$ at the LHC. The
crucial point is that this process has a rather small SM
background. Analysis based on the ATLAS simulation shows the
possibility to detect $\Phi_a^{++}$ as heavy as 1\,TeV if $v_{L}
\sim 29$\,GeV~\cite{Azuelos:2004dm} at LHC. If on the other hand
$v_{L} \simeq v' \ll v_5 \simeq v_{45}$, then the DY production
would dominate and subsequent $\Phi_a^{++} \Phi_a^{--}$ decay into
charged leptons would constitute a clean signal. The most recent
analysis put the LHC reach at around 700\,GeV in the $l^\pm l^\pm$
channel~\cite{Hektor:2007uu,Han:2007bk}.

In our model it is also possible to correlate $\Phi_a^{++}$
detection with the expected proton decay signatures. The main
difference between the two distinct regions in parameter space is
in the strength of the Majorana neutrino Yukawa couplings in
$Y_3$. In the first case these would be extremely small and would
not allow the mapping of the neutrino mass matrix through the
decay of $\Phi_a^{++}$ into a pair of charged leptons:
$\Gamma(\Phi_a^{++} \rightarrow l^+_i l^+_j) \sim
|(Y_{3})_{ij}|^2$. The second case is more promising in that
respect since the relevant Yukawa couplings could be sufficiently
large. In addition, the branching ratios could shed light on the
particular realization of the mass hierarchy in the neutrino
sector, For example, the normal hierarchy scenario implies
$BR(\Phi_a^{++} \rightarrow l^+_i l^+_j) \approx 1/3$ for
$i,j=2,3$.

The best current limits on the $\Phi_a^{\pm \pm}$ mass come from
searches performed at the Tevatron. The lower bound on $\Phi_a^{\pm
\pm}$ comes out to be around 130\,GeV assuming exclusive same-sign
dilepton decays~\cite{Acosta:2004uj}.  This bound however is
derived by explicitly assuming negligible $v_{L}$. In case of
inclusive searches for dilepton events there exists some excess of
events in a recently published analysis~\cite{Abulencia:2007rd}.

To summarize, the main difference between our model and the
majority of models that incorporate a $Y=2$ triplet only lies in (i)
the possibility to have its couplings to gauge bosons
significantly enhanced and (ii) the ability to correlate proton
decay with the triplet detection.

\section{Colour octet}
\label{octet}

Gauge coupling unification constraints impose a firm upper bound on
$M_{\Delta_1}$ in our model. We find that
$M_{\Delta_1} < 250$\,TeV holds for any successful unification
scenario. This is yet another important prediction of our model.
With that in mind we should stress that $\Delta_1$, being an
$SU(3)$ octet that has doublet like couplings to matter, is
phenomenologically very interesting. Its experimental signatures
and relevant limits on its couplings to matter have been recently
discussed within the context of minimal flavour
violation~\cite{Chivukula:1987py,D'Ambrosio:2002ex}. In that
context it is assumed that its couplings to matter and the
corresponding mass matrices of the matter fields are proportional
to each other in the mass eigenstate basis. The phenomenology in that
scenario, the relevant constraints on the
octet couplings and mass as well as a recent analysis of its
potential production at LHC can be found in
Refs.~\cite{Manohar:2006ga,Gresham:2007ri}.

In our case however the couplings of the octet to the matter
fields make only one part of the linear combination that enters
the relevant mass matrices as shown in Eqs.~\eqref{M}. Thus, they
cannot be brought to diagonal form through the same bi-unitary
transformations that define the matter field mass eigenstate
basis. Clearly, the strength of the exchange of neutral components
of $\Delta_1$ will be constrained due to the tree-level
contributions towards $F^0$--$\bar{F}^0$ mixing processes
($F=K,B,D$). For example, using the vacuum saturation
approximation for the hadronic matrix element \cite{Atwood:1996vj}, we
find a new contribution towards $\epsilon_K$ coming from the
$\Delta_1$ exchange to be
\begin{equation}
\label{deltaepsilon} \epsilon_K \simeq \frac{\sqrt{2} f^2_K M_K
B_K}{9 \Delta M_K M^2_{\Delta_1}} \mbox{Im} \left[4(D^T Y_2
D_C)_{21} (D^T Y_2 D_C)^*_{12}\right].
\end{equation}
Using $B_K = 0.75$, $\Delta M_K \simeq 3.48 \times 10^{-12}
$\,MeV, $f_K \simeq 160$\,MeV, $M_K \simeq 498$\,MeV and requiring
that ${\Delta_1}$ exchange contributes to $\epsilon_K$ an amount
less than the experimental value of that quantity
($|\epsilon_K|=2.23 \times 10^{-3}$~\cite{PDG}) gives the
following limit
\begin{equation}
\label{epsilonK} M^2_{\Delta_1} > 2 \times 10^{14} \mbox{Im}
\left[4(D^T Y_2 D_C)_{21} (D^T Y_2
D_C)^*_{12}\right]\,\textrm{GeV}^2.
\end{equation}
Our discussion has made it clear that at least some of the entries
of $Y_2$ have to be non-zero in order to correct equality of down
quark to charged lepton ratios. So, if and when the mass of
$\Delta_1$ is determined, we would have a handle on the strength
of Yukawa couplings of the $\bm{45}$ of Higgs using constrains
such as the one from the $K$ sector. Notice that, unlike a flavour
changing neutral current generating doublets of $SU(2)$ that are
singlets of $SU(3)$ that can also contribute to the processes such
as $\mu$--$e$ conversion and/or $\mu \rightarrow e \gamma$, our
octet is very selective since it couples only to quarks.

\section{$SU(5)$ model comparisons}
\label{compare}
It might come as a surprise that our model with so many Higgs
multiplets does so well in terms of its potential accelerator and
proton decay signatures. To better show the origin and quality of
its predictive power we compare our model with other possible
extensions of the original Georgi-Glashow (GG) $SU(5)$ scenario.

\subsection{$SU(5)$ model with type I see-saw mechanism}
We start with the $SU(5)$ model that extends the GG model with a
$45$ dimensional Higgs representation and at least two
right-handed neutrinos, singlets of $SU(5)$, in order to give
neutrinos their mass. That setup has fewer fields that can
influence gauge coupling unification than our model. It
is in fact already ruled out experimentally unless there exists
some suppression of $d=6$ proton decay operators due to scalar
exchange~\cite{Dorsner:2006dj}. This runs against conclusions
reached in previous analysis~\cite{Giveon:1991zm,Babu:1984vx}.

The only relevant degrees of freedom in that model, as far as the
upper bound on the GUT scale is concerned, are $\Sigma_3$,
$\Delta_1$, $\Delta_3$ and $\Delta_7$. See Table~\ref{tab:table1}
for relevant $b_{ij}$ coefficients. We can thus plot the lines of
constant $M_{\Delta_1}$ and $M_{\Delta_3}$ in the
$M_{GUT}$--$M_{\Sigma_3}$ plane using Eqs.~\eqref{condition1} and
\eqref{condition2} for a fixed mass of $\Delta_7$. In other words,
we can show all viable particle mass spectra of the theory that
yield gauge coupling unification. Recall, $\Delta_7$ is the usual
$SU(2)$ doublet that resides in $\bm{45}$. We set
$M_{\Delta_7}=M_Z$ in order to get the most conservative upper
bound on $M_{GUT}$.

\begin{figure}[ht]
\begin{center}
\includegraphics[width=4in]{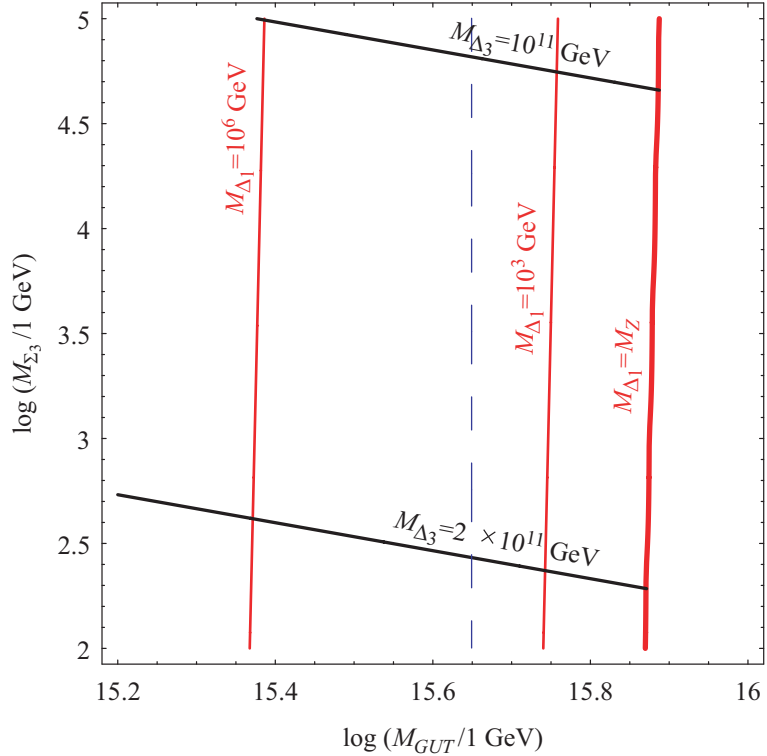}
\end{center}
\caption{\label{figure:one} Particle spectrum that yields exact
gauge coupling unification at the one-loop level for central
values of low-energy observables. A dashed line is the lower bound
on $M_{GUT}$ due to the $d=6$ vector gauge boson mediated proton
decay. The upper bound on $M_{GUT}$ is set by a $M_{\Delta_1}=M_Z$
line. We fix $M_{\Delta_7}=M_Z$ and keep all other fields with
positive $B_{12}$ contributions at the GUT scale.}
\end{figure}

The available parameter space of the theory is shown in
Fig.~\ref{figure:one}. Again, any given point that is not excluded
in Fig.~\ref{figure:one} represents a particle spectrum that
yields exact one-loop unification. A dashed line corresponds to a
lower phenomenological bound on the GUT scale as given by the
proton decay constraints. Clearly, the GUT scale is also bounded
from above at around $10^{15.9}$\,GeV due to the constraint
$M_{\Delta_1} \geq M_Z$. Although unification does take place,
proton decay constraints are not all successfully satisfied. In
particular, the mass of $\Delta_3$, i.e. the scalar that mediates
proton exchange, is below the experimentally inferred bound of
$10^{12}$\,GeV in the otherwise allowed region.
Simply put, the simplest $SU(5)$ model with type I see-saw is
already experimentally ruled out under rather reasonable
assumptions about its flavour structure. We also estimate from
Fig.~\ref{figure:one} that the partial proton decay lifetime due
to the vector gauge boson mediation is at most a factor of $10$
away from its present bound. So, even if the scalar contributions
are assumed to be suppressed, this model would be ruled out by the
next generation of proton decay experiments. In addition, it is
clear that the mass of $\Delta_1$ is bounded from above by the
proton constraints to be less than $10^4$\,GeV.

It is now clear why our model fairs so well. The only additional
fields that affect the GUT scale in our case are $\Phi_a$ and
$\Phi_b$. The former has very small $B_{12}$ contribution and
thus makes no significant disturbance of the unification picture
we present in Fig.~\ref{figure:one}. The latter, which does have a
potentially significant $B_{12}$ contribution, cannot contribute
too much due to the existing lower bound on its mass that
originates from proton decay constraints. If combined, both of
these contributions have just enough strength to satisfy all
phenomenological constraints and allow $M_{\Delta_3}$ at or above
$10^{12}$\,GeV. Clearly, even after the $\bm{15}$ of Higgs is
taken into account, the masses of $\Sigma_3$, $\Delta_1$ and
$\Delta_7$ must still be very low, of the order of the electroweak
scale. That is exactly what we have observed in our previous
discussion of the model with both $\bm{15}$ and $\bm{45}$.
\subsection{$SU(5)$ with the hybrid---type I $+$ type III---see-saw mechanism}
Another model we would like to compare our model with is a recent
extension of the GG model that realizes hybrid see-saw of the type
I and type III nature~\cite{Bajc:2006ia}. (For detailed studies of
that model see Refs.~\cite{Dorsner:2006fx,Bajc:2007zf}.) This
model extends the GG model with only one extra adjoint
representation of fermions $\bm{24}_F
=(\Omega_8,\Omega_3,\Omega_{(3,2)},\Omega_{(\bar{3},
2)},\Omega_{24})=(\bm{8},\bm{1},0)\oplus(\bm{1},\bm{3},0)\oplus(\bm{3},\bm{2},-5/6)
\oplus(\overline{\bm{3}},\bm{2},5/6)\oplus(\bm{1},\bm{1},0)$ and
predicts a very light $SU(2)$ fermionic triplet with $Y=0$. We
first compare these two models on general grounds in terms of
their predictions for the GUT scale. $B_{ij}$ contributions of
$\bm{24}_F$ components are four times larger than those of
corresponding $\Sigma$ components. Hence, in this particular case
$B^{\textrm{min}}_{12}=22/3-1/15-5/15-20/15$ where the third
(fourth) term is due to the $\Sigma_3$ ($\Omega_3$) contribution.
This should be compared with
$B^{\textrm{min}}_{12}=22/3-1/15-5/15-8/15-1/15-1/15-27/15
r_{\Delta_3}-7/15 r_{\Phi_b}$, where $r_{\Phi_b},r_{\Delta_3} \leq
1/3$ due to phenomenological constraints. We thus obtain
comparable values for $B^{\textrm{min}}_{12}$ in both cases. So
even though there are only two degrees of freedom that can
minimize $B_{12}$ in the model with $\bm{24}_F$, their impact on
the running of gauge couplings and hence proton decay predictions
equals the impact of all the fields in our scenario. The important
difference, of course, is that this model is based on
higher-dimensional operators while our model is renormalizable. If
the idea of hybrid see-saw is implemented within the simplest
renormalizable scenario its predictive power is significantly
compromised~\cite{Perez:2007rm}.

Since we are interested in the possibility to test the underlying
mechanism for neutrino mass generation within the grand unified
framework, we assume that the relevant scale for the fields that
generate neutrino mass in both models is $300$\,GeV, i.e.,
$M_{\Omega_3}=M_{\Phi_a}=300$\,GeV, and compare them after we
obtain the upper bound on the GUT scale. The result of this simple
numerical comparison is summarized in Table~\ref{tab:table2}.
\begin{table}[ht]
\caption{\label{tab:table2} Comparison between our model with type
II see-saw mechanism and the hybrid scenario where both type I and
type III see-saw mechanisms are used. We assume
$M_{\Omega_3}=M_{\Phi_a}=300$\,GeV and maximize the GUT scale. In
both cases $M_{\Sigma_3}=M_Z$ and $M_{\Psi_T}=M_{GUT}$.}
\begin{ruledtabular}
\begin{tabular}{lcccccc}
MODEL &$A_{S\,R}$&$A_{S\,L}$& $(M_{GUT}/10^{16}$\,GeV) &
$\alpha^{-1}_{GUT}$ & $\tau^{d=6\textrm{
gauge}}/\tau^{\textrm{exp.}}$
& $\tau^{d=6\textrm{ scalar}}/\tau^{\textrm{exp.}}$ \\
\hline
Dor\v{s}ner-Mocioiu & $2.8$ & $3.0$ & $1.4$  & $29.4$ & $51$& $1$ \\
Bajc-Senjanovi\'{c} & $2.5$ & $2.7$ & $1.5$  & $37.6$ & $150$ & $15000$ \\
\end{tabular}
\end{ruledtabular}
\end{table}

It is evident from Table~\ref{tab:table2} that our model insures
correlations between the direct detection of the field responsible
for the neutrino mass generation and observable proton decay
signatures. That possibility is less likely in the model with the
hybrid see-saw implementation.

\section{Conclusions}
\label{conclusions}
We have investigated a well-motivated $SU(5)$ model which
implements a type II see-saw mechanism for neutrino mass
generation. The model is uniquely determined by requiring
renormalizability, the lack of any additional matter fields
besides those already observed, gauge coupling unification and a
viable phenomenology.

We have shown it is possible to test the underlying mechanism for
neutrino mass generation through accelerator signatures and
correlations with observable proton decay.  The model predicts
that all fields that can participate in electroweak symmetry
breaking are light. Due to a built-in custodial symmetry, the
constraints from precision electroweak measurements are relaxed
compared to standard general analysis and our  model allows a
possible enhancement of the couplings of the $Y=2$ $SU(2)$ triplet
to gauge bosons. This sort of setup can work in any $SU(5)$ theory
with type II see-saw neutrino mass generation. The doubly charged
Higgs present in the model offers promising opportunities for
collider searches. In addition our model predicts a very light
$SU(2)$ doublet that transforms as an octet of $SU(3)$, with
interesting phenomenological consequences. We have also shown that
the proton decay signal is within reach of the next generation of
experiments and it is correlated with the possible collider
signatures of the electroweak scalars. We have also compared our
model with the $SU(5)$ models that implement (i) the type I
see-saw mechanism and (ii) the so-called hybrid scenario that
combines the type I and type III see-saw.  We came to the
conclusion that the minimal $SU(5)$ theory with type I see-saw is
already excluded by experimental limits on partial proton decay
lifetimes. Our model also gives more promising signatures than the
hybrid scenario.

\begin{acknowledgments} This work was supported in part by NSF grant
PHY-0555368. I.M. would like to thank the Aspen Center for Physics
for hospitality while part of this work was completed.
\end{acknowledgments}

\end{document}